\documentclass[11pt]{article}
\usepackage[textwidth=15.2cm,textheight=22cm]{geometry}
\usepackage{amsmath,amssymb}
\usepackage{latexsym}
\usepackage{multicol}
\usepackage{graphicx}
\usepackage{bm}
\usepackage{varwidth}
\usepackage{pifont}
\usepackage{cancel}
\usepackage[dvipsnames]{xcolor}

\allowdisplaybreaks[1]

\newcommand{\del}{\partial}
\newcommand{\be}{\begin{equation}}
\newcommand{\ee}{\end{equation}}
\newcommand{\ba}{\begin{eqnarray}}
\newcommand{\ea}{\end{eqnarray}}
\newcommand{\bdm}{\begin{displaymath}}
\newcommand{\edm}{\end{displaymath}}

\newcommand\fr[1]{\frac{1}{#1}}

\newcommand{\rom}[1]{\uppercase\expandafter{\romannumeral #1\relax}}

\def\ba{\bar A}

\def\beq{\begin{equation}}
\def\eeq{\end{equation}}

\newcommand{\nn}{\nonumber}

\newcommand{\ndt}{\noindent}

\def\bea{\begin{eqnarray}}
\def\eea{\end{eqnarray}}
\def\beas{\begin{eqnarray*}}
\def\eeas{\end{eqnarray*}}
\def\sla{\raise.15ex\hbox{$/$}\kern-.57em}

\def\parm{{\partial}_{-}}

\def\spa#1.#2{\left\langle#1\,#2\right\rangle}
\def\spb#1.#2{\left[#1\,#2\right]}

\begin{document}

\begin{titlepage}
\begin{flushright}    
{\small $\,$}
\end{flushright}
\vskip 1cm
\centerline{\Large{\bf{Higher spins, quadratic forms and amplitudes}}}
\vskip 1cm
\centerline{Sudarshan Ananth, Chetan Pandey and Saurabh Pant}
\vskip 0.3cm
\centerline{\it {Indian Institute of Science Education and Research}}
\centerline{\it {Pune 411008, India}}
\vskip 1.5cm
\centerline{\bf {Abstract}}
\vskip .5cm
The light-cone Hamiltonians for spin 1 and spin 2 fields, describing both the pure and the maximally supersymmetric theories, may be expressed as quadratic forms. In this paper, we show that this feature extends to light-cone higher spin theories. To first order in the coupling constant, we prove that the higher spin Hamiltonians, with and without supersymmetry, are quadratic forms. Scattering amplitude structures emerge naturally in this framework and we relate the momentum space vertex in a supersymmetric higher spin theory to the corresponding vertex in the N=4 Yang-Mills theory.
\vfill
\end{titlepage}

\section{Introduction}

\ndt An interesting feature common to both Yang-Mills theory and gravity is that their light-cone Hamiltonians can  be expressed as quadratic forms~\cite{SA1}. This quadratic form structure appears exclusively in the pure and the maximally supersymmetric varieties ($\mathcal N=4$ superYang-Mills theory and $\mathcal N=8$ Supergravity\footnote{Even in the supersymmetric cases, this quadratic form is {\it {not}} a direct consequence of the fact that the Hamiltonian is the anti-commutator of the supersymmetry generators (see subsection 3.1.1).}). Simple structures like quadratic forms are interesting because they often signal the presence of a hidden symmetry or (and) produce considerable mathematical simplifications in the way we formulate these theories. This is in keeping with evidence that these theories - pure gravity and $\mathcal N=8$ supergravity for example - may possess hidden symmetries in four dimensions~\cite{SA2}. 
\vskip 0.3cm
\ndt This paper focuses entirely on the light-cone Hamiltonian describing interacting higher spin fields~\cite{BBB1,BBB2}, to first order in the coupling constant. Both the non-supersymmetric and supersymmetric theories, in four spacetime dimensions, are examined. We present the following three new results: (1) The light-cone Hamiltonian for higher spin fields without supersymmetry is a quadratic form, (2) Maximally supersymmetric higher spin theories also exhibit this quadratic form structure and (3) The momentum space vertex in a maximally supersymmetric higher spin theory is simply the $\mathcal N=4$ superYang-Mills theory vertex raised to the appropriate power.
\vskip 0.3cm
\ndt The light-cone gauge has two key properties that make it particularly useful when studying scattering amplitude structures. First, it is not manifestly covariant - it has become increasingly clear that preserving manifest covariance obscures much of the simplicity we have come to associate with scattering amplitudes~\cite{HE}. Second, the light-cone gauge focuses exclusively on physical degrees of freedom ensuring that spurious degrees of freedom do not hide the symmetries in a theory. These simplifications in amplitude structures are presented in Section 4. 

\vskip 0.3cm

\section{Cubic interaction vertices in higher spin theories}
We define light-cone co-ordinates in $(-,+,+,+)$ Minkowski space-time as
\begin{eqnarray}
x^{\pm}=\frac{x^{0}\pm x^{3}}{\sqrt{2}} \;,\qquad
x = \frac{x^{1}+ix^{2}}{\sqrt{2}} \;,\qquad\bar{x}= \frac{x^{1}-ix^{2}}{\sqrt{2}}\ .
\end{eqnarray}
The corresponding derivatives are $\partial_{\pm}\,,\,\,\bar{\partial}$ and $\partial$. In four spacetime dimensions, all massless fields have two physical degrees of freedom $\phi$ and $\bar{\phi}$. $\lambda$ is the helicity of the field and $\partial_{+}=\frac{\partial\bar{\partial}}{\partial_{-}}$ for a free theory (modified by corrections when interactions are switched on). $\frac{1}{\parm}$ is defined following the prescription in~\cite{SM}
\vskip 0.3cm
\ndt The Hamiltonian for the free field theory is
\begin{equation}
H\equiv\int d^{3}x\,\mathcal{H}=-\int d^3x\,\bar\phi\,\partial\bar\partial\,\phi\  ,
\end{equation}
with the second equality being valid only for the free theory. We also write
\begin{equation}
\label{hamil}
H\equiv\int d^{3}x\,\mathcal{H}=\int d^{3}x\,\partial_{-}\bar{\phi}\,\delta_{p^-}\phi\ ,
\end{equation}
in terms of the time translation operator
\begin{eqnarray}
\delta_{p^-}\phi \equiv\partial_{+}\phi=\lbrace \phi,\mathcal{H}\rbrace\ .
\end{eqnarray}
In the interacting case, $\delta_{p^-}$ picks up corrections order by order in the coupling constant $g$. 
\vskip 0.3cm
\ndt Details regarding the derivation of light-cone cubic interaction vertices for higher spin theories are presented in~\cite{BBB1,lit}. The idea is to demand closure of the Poincar\'e algebra which restricts and ultimately determines the light-cone Hamiltonian. The result is
\begin{eqnarray}
\delta_{p^-}^{g}\phi = g \sum^{\lambda}_{n=0} (-1)^{n}{\lambda \choose n}\partial_-^{\lambda -1}\left[\frac{\bar{\partial}^{\lambda -n}}{\partial_-^{\,\lambda -n}}\phi\, \frac{\bar{\partial}^{n}}{\partial_-^{\,n}}\phi\right]\ ,
\end{eqnarray}
for even $\lambda$. For odd $\lambda$, algebra-closure requires an antisymmetric structure constant~\cite{SA3}
\begin{equation}
\delta_{p^-}^{g}\phi^a = g f^{abc}\sum^{\lambda}_{n=0} (-1)^{n}{\lambda \choose n}\partial_-^{\lambda -1}\left[\frac{\bar\partial^{\lambda -n}}{\partial_-^{\,\lambda -n}}\phi^b\, \frac{\bar{\partial}^{n}}{\partial_-^{\,n}}\phi^{c}\right]\ .
\end{equation}
From (\ref {hamil}), the complete Hamiltonian to this order reads~\cite{BBB1}
\be
\label{even}
H=\int d^{3}x \left( \partial\bar{\phi}\bar\partial\phi-g \sum^{\lambda}_{n=0} (-1)^{n}{\lambda \choose n}\bar\phi\;\partial_-^\lambda\left[\frac{\bar\partial^{\lambda -n}}{\partial_-^{\,\lambda -n}}\phi \,\frac{\bar{\partial}^{n}}{\partial_-^{\,n}}\phi\right]+c.c. \right),
\ee
for even $\lambda$ and
\be
\label{odd}
H=\int d^{3}x  \left(\partial \bar{\phi}^{a}\bar{\partial}\phi^a-g f^{abc}\sum^{\lambda}_{n=0} (-1)^{n}{\lambda \choose n}\bar\phi^a\;\partial_-^\lambda\left[\frac{\bar\partial^{\lambda -n}}{\partial_-^{\,\lambda -n}}\phi^b\, \frac{\bar{\partial}^{n}}{\partial_-^{\,n}}\phi^{c}\right]+c.c.\right) ,
\ee
for odd $\lambda$. 
\vskip 0.3cm

\subsection{Quadratic forms in higher spin theories - {\bf {without}} supersymmetry}

\ndt In this section, we prove our first claim: that the Hamiltonians in (\ref {even}) and (\ref {odd}) may be written as quadratic forms. Specifically, this means that the Hamiltonians have the following ``whole square" form
\bea
\label{hnonsusy}
\displaystyle H= \int d^3x \,\,\bar{\mathcal D}\phi\,{\mathcal D}\bar\phi\ ,
\eea
in terms of ``covariant'' derivatives. Covariance, specifically for $\lambda=1, 2$ was demonstrated in~\cite{SA1}. We find (structure constants not shown explicitly)
\bea
\displaystyle \mathcal{D}\bar{\phi}=\partial\bar{\phi}-2g\sum_{n=0}^{\lambda-1} (-1)^n{\lambda-1 \choose n} \frac{\bar{\partial}^{n}}{\partial_-^{n+1}} \left[\frac{\bar{\partial}^{\lambda-n-1}}{\partial_-^{\lambda-n-1}} \phi\,\partial_-^\lambda \bar\phi \right]\ ,
\eea 
\bea
\displaystyle \bar{\mathcal{D}}{\phi}=\bar{\partial}\phi-2g\sum_{n=0}^{\lambda-1} (-1)^n{\lambda-1 \choose n} \frac{{\partial}^{n}}{\partial_-^{n+1}} \left[\frac{{\partial}^{\lambda-n-1}}{\partial_-^{\lambda-n-1}} \bar\phi\,\partial_-^\lambda \phi \right]\ ,
\eea 
which reproduce the correct results for both Yang-Mills theory and gravity.
\vskip 0.3cm
\ndt From these definitions, it follows that (\ref {hnonsusy}) correctly produces the kinetic term in the Hamiltonians. To prove its equivalence to (\ref {even}) and (\ref {odd}), we therefore focus on the $O(g)$ contributions from (\ref {hnonsusy}). These are (measure not shown explicitly)
\bea  
-2 g\sum_{n=0}^{\lambda-1} (-1)^n{\lambda-1 \choose n}\left[ \bar{\partial}\phi\,\frac{\bar\partial^n}{\partial_-^{n+1}} \left(\frac{\bar\partial^{\lambda-n-1}}{\partial_-^{\lambda-n-1}} \phi\, \partial_-^\lambda \bar\phi\right) \right]\ ,
\eea 
and its complex conjugate. We partially integrate this expression to obtain
\bea
\label{tyo}
-2g\sum_{n=0}^{\lambda-1} (-1)^{\lambda+n+1}{\lambda-1 \choose n}\, \bar{\phi}\,\partial_-^\lambda\left[\frac{\bar{\partial}^{n+1}}{\partial_-^{n+1}}\phi\frac{\bar{\partial}^{\lambda-n-1}}{\partial_-^{\lambda-n-1}}\phi \right]\ .
\eea 
\ndt We split (\ref{tyo}) into two halves $P$ and $Q$. In $P$, we shift  $n\rightarrow\lambda-n-1$  and invoke the identity
\bea
\label{ci1}
\displaystyle {\lambda-1 \choose n}={\lambda-1 \choose \lambda-1-n}\ ,
\eea
\ndt which yields
\bea
\displaystyle &&P=- g\sum_{n=0}^{\lambda-1} (-1)^n{\lambda-1 \choose n}\,\bar{\phi}\, {\partial_-}^\lambda\left[\frac{\bar{\partial}^{\lambda-n}}{\partial_-^{\lambda-n}}\phi\frac{\bar{\partial}^{n}}{\partial_-^{n}} \phi \right]\ .
\eea
\ndt In the other half $Q$ we shift $n\rightarrow n-1$ to obtain
\bea
&&Q= -g\sum_{n=1}^{\lambda} (-1)^{\lambda+n}{\lambda-1 \choose n-1}\,\bar\phi\,\partial_-^\lambda\left[\frac{\bar\partial^{n}}{\partial_-^n}\phi\frac{\bar\partial^{\lambda-n}}{\partial_-^{\lambda-n}} \phi \right]\ .
\eea 
\ndt We then have
\bea
\displaystyle H=P+Q={\biggl \{}\!\!\!\!\!\!&&- g\sum_{n=0}^{\lambda-1} (-1)^{n}{\lambda-1 \choose n}\, \bar{\phi}\,{\partial_-}^\lambda\left[\frac{\bar\partial^{\lambda-n}}{\partial_-^{\lambda-n}}\phi\frac{\bar\partial^{n}}{\partial_-^{n}} \phi \right]\nn\\
&&-g\sum_{n=1}^{\lambda} (-1)^{\lambda+n}{\lambda-1 \choose n-1}\, \bar\phi\,\partial_-^\lambda\left[\frac{\bar\partial^n}{\partial_-^n}\phi\frac{\bar\partial^{\lambda-n}}{\partial_-^{\lambda-n}} \phi\right] {\biggl \}}\ ,\nn \\
=&&-\displaystyle g\sum_{n=0}^{\lambda} (-1)^n \left(\left[{\lambda-1 \choose n}+{\lambda-1 \choose n-1} \right] \bar{\phi}\, \partial_-^\lambda\left[\frac{\bar\partial^{\lambda-n}}{\partial_-^{\lambda-n}}\phi\frac{\bar\partial^{n}}{\partial_-^{n}} \phi \right]\right)\ .
\eea
Using the Pascal triangle property
\bea
\label{pascal}
{\lambda-1 \choose n}+{\lambda-1 \choose n-1}={\lambda \choose n}\ ,
\eea
\ndt this is
\bea
\displaystyle H=P+Q=-\displaystyle g\sum_{n=0}^{\lambda} (-1)^n {\lambda \choose n}\, \bar\phi\,\partial_-^\lambda\left[\frac{\bar\partial^{\lambda-n}}{\partial_-^{\lambda-n}}\phi\frac{\bar\partial^n}{\partial_-^n} \phi \right]\ ,
\eea 
reproducing the structures in (\ref {even}), (\ref {odd}) and confirming that the higher spin Hamiltonians are indeed quadratic forms.

\vskip 0.3cm

\section{Cubic interaction vertices in supersymmetric higher spin theories}

\ndt We now move to theories of arbitrary integer spin {\it {with}} supersymmetry. We work in light-cone superspace where the supersymmetry generators are of two varieties~\cite{BBB2}. Realized on Grassmann parameters $\theta^m$ and their conjugates $\bar\theta_m$, the kinematical generators are
\bea
q^{m}=-\frac{\partial}{\partial \bar{\theta}_{m}}-\frac{i}{\sqrt{2}} \theta^{m} \partial_{-}  \;,\;\;\;\;\;\;\;\; \bar{q}_{n}=\frac{\partial}{\partial \theta^{n}}+\frac{i}{\sqrt{2}} \bar{\theta}_{n} \partial_{-}\ ,
\eea 
and satisfy
\bea
\left \{ q^m, \bar{q}_{n} \right \}=-\sqrt{2}\delta^m_np^+\ .
\eea 
\ndt The dynamical generators satisfy
\bea
\{Q^m,{\bar Q}_n\}=\delta^m_n\,p^-\ ,
\eea
thus representing ``square roots'' of the light-cone Hamiltonian. At lowest order, they are
\bea
Q^m=-\frac{\bar\partial}{\partial_-}\,q^m\ , \qquad {\bar Q}_m=-\frac{\partial}{\partial_-}\,{\bar q}_m\ .
\eea
These expression pick up corrections order by order in the coupling constant in an interacting theory. We introduce superspace derivatives $d^m$ and $\bar{d_n}$ 
\begin{eqnarray}
d^{m}=-\frac{\partial}{\partial \bar{\theta}_{m}}+\frac{i}{\sqrt{2}} \theta^{m} \partial_{-}  \;, \;\;\;\;\;\;\;\; {\bar{d}}_{n}=\frac{\partial}{\partial \theta^{n}}-\frac{i}{\sqrt{2}} \bar{\theta}_{n} \partial_{-}\ ,
\end{eqnarray} 
\ndt which anti-commute with the supersymmetry generators and satisfy
\bea
\label{cd1}
\left\{d^m, \bar{d_n} \right \} =i\sqrt{2}\delta_n^{m}\partial_-\ .
\eea
\ndt We use $\Phi$ to denote a superfield and impose a ``chirality" condition on it so
\bea
\displaystyle \mathrm{d}^m \Phi (x, \theta,\bar{\theta})=0\ .
\eea 
\ndt For maximally extended supermultiplets, we have the additional ``inside-out" constraint
\bea
\label{io}
\bar{\Phi}(x, \theta, \bar{\theta})=\frac{1}{2^{N / 4}N !} \frac{\bar{\mathrm{d}}^{N}}{\partial_-^{\,N / 2}} \Phi(x, \theta, \bar{\theta})\ ,
\eea
with $N=4\,\lambda$. The Hamiltonian in this formalism is 
\begin{equation}
H\equiv\int d^{3}x\,\mathrm{d}^{N} \theta\, \mathrm{d}^{N}\bar{\theta}\;\mathcal{H}\;\;=\;\int d^{3}x\,\mathrm{d}^{N} \theta \,\mathrm{d}^{N}\bar{\theta}\;\partial_{-}\bar\Phi\,\delta_{p^{-}}\Phi\ .
\end{equation}
Algebra-closure now involves the larger superPoincar\'e algebra~\cite{SA4} and yields the following results~\cite{BBB2}
\ndt
\bea
\label{queue}
Q^m\,\bar\Phi\,=-\displaystyle \frac{\bar{\partial}}{\partial_{-}} q^{m} \bar{\Phi}-2g\sum_{n=0}^{\lambda-1} (-1)^n{\lambda-1 \choose n} \frac{1}{\partial_{-}}\left(\partial^{\left(\lambda-1-n \right)}\partial_-^n \; d^{m} \bar{\Phi}\; \partial_{-}^{\left(\lambda-n \right)} \partial^n \bar{\Phi} \right)\!+\!\mathcal{O}\left(g^{2}\right)
\eea
for the dynamical supersymmetry generator
\bea
\delta_{p^{-}}^{g} \Phi=g\sum_{n=0}^{\lambda} (-1)^n{\lambda \choose n} \frac{1}{\partial_{-}}\left[\bar{\partial}^{(\lambda-n)} \partial_-^{\,n} \Phi \; \bar{\partial}^{n} \partial_-^{\,(\lambda-n)} \Phi\right]\ ,
\eea
\ndt for even $\lambda$ and
\bea
\delta_{p^{-}}^{g} \Phi^a=gf^{abc}\sum_{n=0}^{\lambda} (-1)^n{\lambda \choose n} \frac{1}{\partial_{-}}\left[\bar{\partial}^{(\lambda-n)} \partial_-^{\,n} \Phi^b \; \bar{\partial}^{n} \partial_-^{(\lambda-n)} \Phi^c\right]\ ,
\eea
for odd $\lambda$. The corresponding Hamiltonians being
\bea
\label{even1}
\displaystyle H=&&\!\!\!\!\!\!\int {\mathrm d}^3 x\, \mathrm{d}^{N} \theta \,{\mathrm d}^N\bar\theta\;\;{\biggl \{}\frac{1}{2} \partial\bar\Phi \frac{\bar\partial}{\partial_-^{\,N / 2}} \Phi \\
&&\!\!\!\!\!\!-\frac{1}{3}g \left(\frac{1}{\partial_-^{\,N/2}}\bar{\Phi}\sum_{n=0}^{\lambda} (-1)^n{\lambda \choose n} \left[\bar{\partial}^{(\lambda-n)} \partial_-^{\,n} \Phi \; \bar{\partial}^{n} \partial_-^{\,(\lambda-n)} \Phi\right]+c.c. \right)+\mathcal{O}(g^2){\biggl \}}\nn \ ,
\eea
\ndt for even $\lambda$ and
\bea
\label{odd1}
\displaystyle H=&&\!\!\!\!\!\!\int \mathrm{d}^3 x\, \mathrm{d}^N \theta\, \mathrm{d}^N\bar\theta\;\;{\biggl \{}\frac{1}{2} \partial \bar{\Phi}^a \frac{\bar{\partial}}{\partial_-^{\,N / 2}} \Phi^a \\
&&\!\!\!\!\!\!-\frac{1}{3}gf^{abc} \left[\frac{1}{\partial_-^{\,N/2}}\bar{\Phi}^{a}\sum_{n=0}^{\lambda} (-1)^n{\lambda \choose n} \left[\bar{\partial}^{(\lambda-n)} \partial_-^{n} \Phi^b \; \bar{\partial}^{n} \partial_-^{(\lambda-n)} \Phi^c\right]+c.c. \right]+\mathcal{O}(g^2) {\biggl \}}\nn \ ,
\eea
\ndt for odd $\lambda$.

\vskip 1cm

\subsection{Quadratic forms in higher spin theories - {\bf {with}} supersymmetry}

\ndt We now prove, to first order in the coupling constant, that the light-cone Hamiltonians in (\ref {even1}) and (\ref {odd1}) are quadratic forms. The specific claim is that these Hamiltonians can be written as follows.
\bea
\label{hsusy}
H&&\!\!\!\!\!\!=\displaystyle \frac{2}{N\sqrt{2}}  \left(\mathcal{W}_m , \; \mathcal{W}_m \right)\ ,
\eea
\ndt with
\bea
\left(\Phi , \; \Xi \right)\;\equiv\;\displaystyle 2i \int d^{3} x\, d^{N} \theta\, d^{N} \bar{\theta} \; \bar\Phi \fr{\partial_-^{2\lambda-1}} \Xi\ .
\eea
\vskip 1cm

\subsubsection{Quadratic form $\neq$ anti-commutator}

Before proving (\ref {hsusy}), we explain how the ``quadratic form'' concept differs from the idea of writing the Hamiltonian as the anti-commutator of dynamical supersymmetries. This is best illustrated by restricting ourselves to the lowest order  dynamical supersymmetry generators. Start with the simple choice\footnote{This choice preserves chirality, ie. acting with a $\bar d$ on (\ref {dublew}) yields $0$.}
\bea
\label{dublew}
\overline{\mathcal W}^m=Q^m\,\bar\Phi\ ,
\eea
where keep just the $g=0$ piece in (\ref {queue}). The claim in (\ref {hsusy}) is that the Hamiltonian reads (measure, integrals and constants are suppressed and a factor of $2$ included for convenience)
\bea
\label{claim}
 H=\left(\mathcal{W}_m , \; \mathcal{W}_m \right)=2\,Q^m\bar\Phi\,\fr{\partial_-^{2\lambda-1}}\,{\bar Q}_m\,\Phi\ .
\eea
The step above is valid in any theory with supersymmetry. However, exclusive to maximally supersymmetric theories is the inside-out relation in~(\ref {io}) which we now invoke. 
Write (\ref {claim}) as two halves and apply the inside-out relation only to the second expression to obtain
\bea
\label{step}
 H=Q^m\bar\Phi\,\fr{\partial_-^{2\lambda-1}}\,{\bar Q}_m\,\Phi + Q^m\fr{\partial_-^{2\lambda}}\Phi\,\partial_-{\bar Q}_m\,\bar\Phi\ .
\eea
\ndt {\it {Rewriting the second term this way is not possible in theories with less than maximal supersymmetry}}. 
\vskip 0.3cm
\ndt In (\ref {step}), integrate the $Q$ in the first expression to the right and the $\bar Q$ in the second expression to the left to obtain
\bea
\label{step2}
 H=-\bar\Phi\,\fr{\partial_-^{2\lambda-1}}\,Q^m\,{\bar Q}_m\,\Phi + {\bar Q}_m\,Q^m\fr{\partial_-^{2\lambda}}\Phi\,\partial_-\bar\Phi\ .
\eea
Finally, in the second expression, integrate the $\partial_-$ once leaving us with
\bea
 H=-\bar\Phi\,\fr{\partial_-^{2\lambda-1}}\,Q^m\,{\bar Q}_m\,\Phi - {\bar Q}_m\,Q^m\fr{\partial_-^{2\lambda-1}}\Phi\,\bar\Phi\ ,
\eea
which is equivalent to
\bea
\label{max}
 H=-\bar\Phi\,\fr{\partial_-^{2\lambda-1}}\,\{\,Q^m\,,\,{\bar Q}_m\,\}\,\Phi\ ,
\eea
known to be a true statement. Thus a quadratic form structure as in (\ref {hsusy}) can only appear in maximally supersymmetric theories, where it is equivalent to (\ref {max}) at the lowest order.
\vskip 0.3cm
\ndt It is important to remember that the $Q$ in (\ref {step}) is non-linearly realized on the superfields beyond the lowest order and cannot be integrated as in (\ref {step2}) if we were to include higher order contributions to (\ref {dublew}).

\vskip 1cm

\subsubsection{The proof}

\ndt We have already identified $\mathcal{W}$ with the action of the dynamical supersymmetry on the superfield in (\ref{dublew}). Accordingly, including the first order contributions we have
\bea
\displaystyle \overline{\mathcal{W}}^{m}=\displaystyle -\frac{\bar{\partial}}{\partial_{-}} q^{m} \bar{\Phi}-2g\sum_{k=0}^{\lambda-1} (-1)^k{\lambda-1 \choose k} \frac{1}{\partial_{-}}\left(\partial^{\left(\lambda-1-k \right)}{\partial_-}^{\left(k \right)} d^{m} \bar{\Phi}\; \partial_{-}^{\left(\lambda-k \right)} \partial^k \bar{\Phi} \right)+\mathcal{O}\left(g^{2}\right)
\eea
with the appropriate structure constants for odd $\lambda$. The kinetic contribution from (\ref {hsusy}) is trivial. At cubic order, the Hamiltonian involves terms of the form
\bea
\displaystyle H=\frac{8i}{N\sqrt{2}}g \frac{\bar{\partial}}{\partial_-^{2\lambda+1}}q^m \; \bar{\Phi} \sum_{k=0}^{\lambda-1} (-1)^k {\lambda-1 \choose k} \left( \bar{\partial}^{(\lambda-1-k)} \partial_-^k \; \bar{d}_m \; \Phi \; \partial_-^{(\lambda-k)}\bar{\partial}^k \; \Phi \right)\ .
\eea
\ndt We use $q^m X=-i\sqrt{2}\theta^m \partial_- X$ where $X$ is any chiral combination of superfields~\cite{SA4}, and $\theta^m \bar{d}_m=N\, \theta^1 {\bar d}_1$ to simplify this to 
\bea
\label{Cubic}
\displaystyle H= 8g \frac{\bar{\partial}}{\partial_-^{2\lambda}}\theta^1 \; \bar{\Phi} \sum_{k=0}^{\lambda-1}  (-1)^k {\lambda-1 \choose k} \left( \bar{\partial}^{(\lambda-1-k)} \partial_-^k \; \bar{d}_1 \; \Phi \; \partial_-^{(\lambda-k)}\bar{\partial}^k \; \Phi \right)\ .
\eea
\ndt Equation (\ref {Cubic}) is our starting point and we will rewrite it in two different ways. The first rewriting involves integrating the $\bar\partial$ from the first superfield to produce two terms, $I$ and $J$
\bea
\label{IJ}
\displaystyle H=\!\!\!\!\!\!&&\color{PineGreen}{-8g \frac{1}{\partial_-^{2\lambda}}\theta^1 \; \bar{\Phi} \sum_{k=0}^{\lambda-1}  (-1)^k {\lambda-1 \choose k} \left( \bar{\partial}^{(\lambda-k)} \partial_-^k \; \bar{d}_1 \; \Phi \; \partial_-^{(\lambda-k)}\bar{\partial}^k \; \Phi \right)}\nn \\
&&\color{PineGreen}{-8g \frac{1}{\partial_-^{2\lambda}}\theta^1 \; \bar{\Phi} \sum_{k=0}^{\lambda-1}  (-1)^k {\lambda-1 \choose k} \left( \bar{\partial}^{(\lambda-1-k)} \partial_-^k \; \bar{d}_1 \; \Phi \; \partial_-^{(\lambda-k)}\bar{\partial}^{(k+1)} \; \Phi \right)}\nn \ ,\\
=\!\!\!\!\!\!&&I+J\ .
\eea
This form for the Hamiltonian will be used later in this subsection. 
\vskip 0.3cm
\ndt We now rewrite (\ref {Cubic}) in a second manner (terms in {\textcolor{blue} {blue}} survive the manipulations described below). The first step is to partially integrate the $\bar{d}_1$ in (\ref{Cubic}), to obtain two terms
\bea
\displaystyle H=\!\!\!\!\!\!&&\textcolor{blue}{+8g \frac{\bar{\partial}}{\partial_-^{2\lambda}} \; \bar{\Phi}\sum_{k=0}^{\lambda-1}  (-1)^k {\lambda-1 \choose k} \left( \bar{\partial}^{(\lambda-1-k)} \partial_-^k  \; \Phi \; \partial_-^{(\lambda-k)}\bar{\partial}^k \; \Phi \right)}\ ,\\
\label{second}
&&-8g \frac{\bar{\partial}}{\partial_-^{2\lambda}}\; \bar{\Phi}\sum_{k=0}^{\lambda-1}  (-1)^k {\lambda-1 \choose k} \left( \bar{\partial}^{(\lambda-1-k)} \partial_-^k  \; \Phi \; \partial_-^{(\lambda-k)}\bar{\partial}^k \; \theta^1 \bar{d}_1 \; \Phi \right)\ .
\eea
\ndt In (\ref {second}), a $\partial_-$ is integrated away from the last superfield to yield
\bea
\label{X}
\!\!\!\!\!\!&&+8g \frac{\bar{\partial}}{\partial_-^{2\lambda}}\; \bar{\Phi}\sum_{k=0}^{\lambda-1}  (-1)^k {\lambda-1 \choose k} \left( \bar{\partial}^{(\lambda-1-k)} \partial_-^{(k+1)}  \; \Phi \; \partial_-^{(\lambda-k-1)}\bar{\partial}^k \; \theta^1 \bar{d}_1 \; \Phi \right)\ ,\\
\label{G}
&&+8g \frac{\bar{\partial}}{\partial_-^{2\lambda-1}}\; \bar{\Phi}\sum_{k=0}^{\lambda-1}  (-1)^k {\lambda-1 \choose k} \left( \bar{\partial}^{(\lambda-1-k)} \partial_-^{k}  \; \Phi \; \partial_-^{(\lambda-k-1)}\bar{\partial}^k \; \theta^1 \bar{d}_1 \; \Phi \right)\ .
\eea
\ndt Equation (\ref{X}) is the negative of (\ref{Cubic}) since
\beas
\label{ci12}
\displaystyle {\lambda-1 \choose k}={\lambda-1 \choose \lambda-1-k}\ ,
\eeas
allowing us to combine it with (\ref{Cubic}) producing a factor of two. Equation (\ref{G}) can be simplifed, using integrations and the inside-out constraint, to
\bea
\label{H}
&&+4g\; \bar{\partial}\partial_- \;{\Phi} \sum_{k=0}^{\lambda-1}  (-1)^k {\lambda-1 \choose k} \left( \bar{\partial}^{(\lambda-1-k)} \partial_-^{k}\frac{1}{\partial_-^{2\lambda}} \; \bar{\Phi} \; \partial_-^{(\lambda-k-1)}\bar{\partial}^k \; \theta^1 \bar{d}_1 \; \Phi \right)\ ,\\
\label{J}
&&-4g\; \bar{\partial}\partial_- \; \bar{d}_1{\Phi}\sum_{k=0}^{\lambda-1}  (-1)^k {\lambda-1 \choose k} \left( \bar{\partial}^{(\lambda-1-k)} \partial_-^{k}\frac{1}{\partial_-^{2\lambda}} \; \bar{\Phi} \; \partial_-^{(\lambda-k-1)}\bar{\partial}^k \; \theta^1 \; \Phi \right)\ ,\\
\label{K}
&&-4g\; \bar{\partial}\; \bar{d}_1{\Phi}\sum_{k=0}^{\lambda-1}  (-1)^k {\lambda-1 \choose k} \left( \bar{\partial}^{(\lambda-1-k)} \partial_-^{k}\frac{1}{\partial_-^{2\lambda-1}} \; \bar{\Phi} \; \partial_-^{(\lambda-k-1)}\bar{\partial}^k \; \theta^1 \; \Phi \right)\ .
\eea
\ndt We simplify (\ref{J}) by integrating the $\bar{d}_1$ from the first superfield, yielding
\bea
\label{L}
&&\textcolor{blue}{+4g\; \bar{\partial}\partial_- \; {\Phi}\sum_{k=0}^{\lambda-1}  (-1)^k {\lambda-1 \choose k} \left( \bar{\partial}^{(\lambda-1-k)} \partial_-^{k}\frac{1}{\partial_-^{2\lambda}} \; \bar{\Phi} \; \partial_-^{(\lambda-k-1)}\bar{\partial}^k \; \Phi \right)}\ ,\\
\label{H'}
&&-4g\; \bar{\partial}\partial_- \; {\Phi}\sum_{k=0}^{\lambda-1}  (-1)^k {\lambda-1 \choose k} \left( \bar{\partial}^{(\lambda-1-k)} \partial_-^{k}\frac{1}{\partial_-^{2\lambda}} \; \bar{\Phi} \; \partial_-^{(\lambda-k-1)}\bar{\partial}^k \; \theta^1 \bar{d}_1\; \Phi \right)\ ,
\eea
and find that (\ref{H'}) cancels against (\ref{H}). Integrating the chiral derivative and using partial integrations simplifies (\ref{K}) to
\bea
\label{K'}
&&\textcolor{blue}{+4g\; \bar{\partial}\; {\Phi}\sum_{k=0}^{\lambda-1}  (-1)^k {\lambda-1 \choose k} \left( \bar{\partial}^{(\lambda-1-k)} \partial_-^{k}\frac{1}{\partial_-^{2\lambda-1}} \; \bar{\Phi} \; \partial_-^{(\lambda-k-1)}\bar{\partial}^k \;  \Phi \right)}\ ,\\
\label{-G}
&&-4g\;\frac{\bar{\partial}}{\partial_-^{2\lambda-1}} \; \bar{\Phi}\sum_{k=0}^{\lambda-1} (-1)^k {\lambda-1 \choose k}\bar{\partial}^{(\lambda-1-k)}\partial_-^k \left({\Phi} \; \partial_-^{(\lambda-k-1)}\bar{\partial}^k \; \theta^1 \bar{d}_1 \; \Phi \right)\ ,\\
\label{Junk}
&&+4g\;\frac{\bar{\partial}}{\partial_-^{2\lambda-1}} \; \bar{\Phi}\sum_{k=0}^{\lambda-1} (-1)^k {\lambda-1 \choose k}\bar{\partial}^{(\lambda-2-k)}\partial_-^k \left({\Phi} \; \partial_-^{(\lambda-k-1)}\bar{\partial}^{(k+1)} \; \theta^1 \bar{d}_1 \; \Phi \right)\ .
\eea
\ndt Note that (\ref{-G}) is the negative of (\ref{G}) using the identity (\ref {magic1}) from appendix {\bf A} (these terms combine with a factor of two). Partial integrations of $\partial_-$ and $\bar\del$ in (\ref {Junk}) followed by the use of a similar identity (\ref {magic2}) yields
\bea
&&\textcolor{blue}{+2g\;\frac{1}{\partial_-^{2\lambda}} \; \bar{\Phi}\sum_{k=0}^{\lambda-1} (-1)^k {\lambda-1 \choose k} \left(\bar{\partial}^{\lambda-1-k}\partial_-^{k}\; {\Phi}\; \partial_-^{(\lambda-k)}\bar{\partial}^{(k+1)} \; \theta^1 \bar{d}_1 \;\;  \Phi \right)}\ ,\\
&&\textcolor{blue}{+2g\;\frac{1}{\partial_-^{2\lambda}} \; \bar{\Phi}\sum_{k=0}^{\lambda-1} (-1)^k {\lambda-1 \choose k} \left(\bar{\partial}^{\lambda-1-k}\partial_-^{(k+1)}\; {\Phi}\; \partial_-^{(\lambda-k-1)}\bar{\partial}^{(k+1)} \; \theta^1 \bar{d}_1 \;\;  \Phi \right)}\ .
\eea
\ndt This completes the second re-writing of (\ref {Cubic}). 
\vskip 0.3cm
\ndt We have now rewritten (\ref {Cubic}) in two ways. We now subtract $\fr{4}$ times the first form in (\ref {IJ}), written in {\textcolor{PineGreen} {green}}, from the second form of (\ref {Cubic}), in {\textcolor{blue}{blue}}. Thus 
\bea
\label{1}
{H}-\frac{1}{4}{H}=\!\!\!\!\!\!\!\!\!\!&&\textcolor{blue}{+4g \frac{\bar{\partial}}{\partial_-^{2\lambda}} \; \bar{\Phi}\sum_{k=0}^{\lambda-1}  (-1)^k {\lambda-1 \choose k} \left( \bar{\partial}^{(\lambda-1-k)} \partial_-^k  \; \Phi \; \partial_-^{(\lambda-k)}\bar{\partial}^k \; \Phi \right)}\\
\label{2}
&&\textcolor{blue}{+2g\; \bar{\partial}\partial_- \; {\Phi}\sum_{k=0}^{\lambda-1}  (-1)^k {\lambda-1 \choose k} \left( \bar{\partial}^{(\lambda-1-k)} \partial_-^{k}\frac{1}{\partial_-^{2\lambda}} \; \bar{\Phi} \; \partial_-^{(\lambda-k-1)}\bar{\partial}^k \; \Phi \right)}\\
\label{3}
&&\textcolor{blue}{+2g\; \bar{\partial}\; {\Phi}\sum_{k=0}^{\lambda-1}  (-1)^k {\lambda-1 \choose k} \left( \bar{\partial}^{(\lambda-1-k)} \partial_-^{k}\frac{1}{\partial_-^{2\lambda-1}} \; \bar{\Phi} \; \partial_-^{(\lambda-k-1)}\bar{\partial}^k \;  \Phi \right)}\\
\label{4}
&&\textcolor{blue}{+2g\;\frac{1}{\partial_-^{2\lambda}} \; \bar{\Phi}\sum_{k=0}^{\lambda-1} (-1)^k {\lambda-1 \choose k} \left(\bar{\partial}^{\lambda-1-k}\partial_-^{k}\; {\Phi}\; \partial_-^{(\lambda-k)}\bar{\partial}^{(k+1)} \; \theta^1 \bar{d}_1 \;\;  \Phi \right)}\\
\label{5}
&&\color{PineGreen}{-2g \frac{1}{\partial_-^{2\lambda}}\theta^1 \; \bar{\Phi} \sum_{k=0}^{\lambda-1}  (-1)^k {\lambda-1 \choose k} \left( \bar{\partial}^{(\lambda-1-k)} \partial_-^k \; \bar{d}_1 \; \Phi \; \partial_-^{(\lambda-k)}\bar{\partial}^{(k+1)} \; \Phi \right)}\\
\label{6}
&&\color{PineGreen}{+2g \frac{1}{\partial_-^{2\lambda}}\theta^1 \; \bar{\Phi} \sum_{k=0}^{\lambda-1}  (-1)^k {\lambda-1 \choose k} \left( \bar{\partial}^{(\lambda-k)} \partial_-^k \; \bar{d}_1 \; \Phi \; \partial_-^{(\lambda-k)}\bar{\partial}^k \; \Phi \right)}\\
\label{7}
&&\textcolor{blue}{+2g\;\frac{1}{\partial_-^{2\lambda}} \; \bar{\Phi}\sum_{k=0}^{\lambda-1} (-1)^k {\lambda-1 \choose k} \left(\bar{\partial}^{\lambda-1-k}\partial_-^{(k+1)}\; {\Phi}\; \partial_-^{(\lambda-k-1)}\bar{\partial}^{(k+1)} \; \theta^1 \bar{d}_1 \; \Phi \right)}
\eea
Terms (\ref{4}) and (\ref{5}) combine into a single term (chain rule for ${\bar d}_1$). Terms (\ref{6}) and (\ref{7}) cancel due to the combinatorial identity. Integration of $\partial_-$ in (\ref{2}) produces two terms, one of which cancels (\ref{3}). The inside-out relation simplifies the other term to
\bea
&&-2g\; \frac{\bar{\partial}}{\partial_-^{2\lambda}} \; \bar{\Phi}\sum_{k=0}^{\lambda-1}  (-1)^k {\lambda-1 \choose k} \left( \bar{\partial}^{(\lambda-1-k)} \partial_-^{k} \;{\Phi} \; \partial_-^{(\lambda-k)}\bar{\partial}^k \; \Phi \right)\ ,
\eea
which combines with (\ref{1}). In this combination, we integrate the $\bar{\partial}$ to obtain two terms
\bea
\label{Answer}
&&-2g\; \frac{1}{\partial_-^{2\lambda}} \; \bar{\Phi}\sum_{k=0}^{\lambda-1}  (-1)^k {\lambda-1 \choose k} \left( \bar{\partial}^{(\lambda-k)} \partial_-^{k} \;{\Phi} \; \partial_-^{(\lambda-k)}\bar{\partial}^k \; \Phi \right)\ ,\\
\label{Cancel}
&&-2g\; \frac{1}{\partial_-^{2\lambda}} \; \bar{\Phi}\sum_{k=0}^{\lambda-1}  (-1)^k {\lambda-1 \choose k} \left( \bar{\partial}^{(\lambda-1-k)} \partial_-^{k} \;{\Phi} \; \partial_-^{(\lambda-k)}\bar{\partial}^{(k+1)} \; \Phi \right)\ ,
\eea
with (\ref{Cancel}) canceling the (\ref{4})-(\ref{5}) combine leaving us with (\ref{Answer}). We simplify (\ref {Answer}) as follows.
\bea
\displaystyle H\!\!\!\!\!\!&&=-\frac{8}{3}g\; \frac{1}{\partial_-^{2\lambda}} \; \bar{\Phi}\sum_{k=0}^{\lambda-1}  (-1)^k {\lambda-1 \choose k} \left( \bar{\partial}^{(\lambda-k)} \partial_-^{k} \;{\Phi} \; \partial_-^{(\lambda-k)}\bar{\partial}^k \; \Phi \right)\\
&&=-\frac{4}{3}g\; \frac{1}{\partial_-^{2\lambda}} \; \bar{\Phi}\sum_{k=0}^{\lambda}  (-1)^k \left[{\lambda-1 \choose k}+{\lambda-1 \choose k-1} \right] \left( \bar{\partial}^{(\lambda-k)} \partial_-^{k} \;{\Phi} \; \partial_-^{(\lambda-k)}\bar{\partial}^k \; \Phi \right)\\
&&=-\frac{4}{3}g\; \frac{1}{\partial_-^{2\lambda}} \; \bar{\Phi}\sum_{k=0}^{\lambda}  (-1)^k {\lambda \choose k} \left( \bar{\partial}^{(\lambda-k)} \partial_-^{k} \;{\Phi} \; \partial_-^{(\lambda-k)}\bar{\partial}^k \; \Phi \right)\ ,
\eea
matching the structures in (\ref {even1}) and (\ref {odd1}) confirming that these are quadratic forms.

\vskip 0.3cm

\section{Amplitude structures in higher spin theories}

We now examine the scattering amplitude structures that appear in the cubic Hamiltonians discussed thus far. Any four-vector can be expressed as a bispinor using the Pauli matrices, $p_{a\dot{a}}=p_{\mu}\sigma^{\mu}_{a\dot{a}}$, with $\det(p_{a\dot{a}})$ yielding $-p^{\mu}p_{\mu}$. We introduce the spinor product
\begin{equation}
\label{shp}
\spa{k}.{l}\equiv \sqrt{2}\frac{(kl_{-}-lk_{-})}{\sqrt{k_{-}l_{-}}}\ .
\end{equation}

\vskip 0.3cm

\subsection{The non-supersymmetric case}

\ndt Scattering amplitudes in non-supersymmetric higher spin theories, to first order in the coupling, were discussed in~\cite{SA6}. The main result, in momentum space, is that the cubic vertices in (\ref {even}) and (\ref {odd}) may be obtained by raising the cubic vertex in pure Yang-Mills theory (the $\lambda=1$ case), to the appropriate power~\cite{SA7}.
\bea
\label{result}
{\it L^{\,\lambda}_{\,3}}={{\biggl [}\frac{\spa{k}.{l}^{3}}{\spa{l}.{p}\spa{p}.{k}}{\biggr ]}}^\lambda={\biggl [}{\it L^{\,{\mbox {YM}}}_{\,3}}{\biggr ]}^\lambda\  .
\eea
\vskip 0.3cm

\subsection{The supersymmetric case}

We turn now to the third new result in this paper, pertaining to scattering amplitude structures in maximally supersymmetric higher spin theories. The actions corresponding to (\ref {even1}) and (\ref {odd1}) are
\bea
\label{even1S}
\displaystyle S&=&\int \mathrm{d}^{4} x\, \mathrm{d}^{N} \theta \,\mathrm{d}^{N}\bar{\theta}\;\;{\biggl \{}\frac{1}{4} \bar{\Phi} \frac{\square}{\partial_-^{\,N / 2}} \Phi \\
+&&\frac{1}{3}g \left(\frac{1}{\partial_-^{\,N/2}}\Phi\sum_{n=0}^{\lambda} (-1)^n{\lambda \choose n} \left[\partial^{\lambda-n} \partial_-^n\bar\Phi \; \partial^n \partial_-^{\lambda-n} \bar\Phi\right]+c.c. \right)+\mathcal{O}(g^2){\biggl \}}\ , \nn
\eea
\ndt for even $\lambda$ , and
\bea
\label{odd1S}
\displaystyle S&=&\int \mathrm{d}^{4} x\, \mathrm{d}^{N} \theta\, \mathrm{d}^{N}\bar{\theta}\;\;{\biggl \{}\frac{1}{4} \bar{\Phi}^a \frac{\square}{\partial_-^{\,N / 2}} \Phi^a \\
+&&\frac{1}{3}gf^{abc} \left[\frac{1}{\partial_-^{\,N/2}}\Phi^a\sum_{n=0}^{\lambda} (-1)^n{\lambda \choose n} \left[\partial^{\lambda-n} \partial_-^{n} \bar\Phi^b \; \partial^n \partial_-^{\lambda-n} \bar\Phi^c\right]+c.c. \right]+\mathcal{O}(g^2) {\biggl \}}\ , \nn
\eea
\ndt for odd $\lambda$. In momentum space, both cubic vertices have the following basic structure (measure and constants suppressed)
\bea
\displaystyle &&\frac{\delta^4(p+k+l)}{(k_-+l_-)^{2\lambda}}\sum_{n=0}^{\lambda} (-1)^n {\lambda \choose n} \left[k^{\lambda-n}\;k_-^{n}\; l^n\; l_-^{\lambda-n}\right] \tilde{\Phi}(p)\tilde{\bar{\Phi}}(k)\tilde{\bar{\Phi}}(l)+c.c.\nn\\
&&=\frac{\delta^4(p+k+l)}{(k_-+l_-)^{2\lambda}}\sum_{n=0}^{\lambda} (-1)^n {\lambda \choose n} \left[(kl_-)^{\lambda-n}\;(k_-l)^n \right] \tilde{\Phi}(p)\tilde{\bar{\Phi}}(k)\tilde{\bar{\Phi}}(l)+c.c.\nn\\
&&=\delta^4(p+k+l)\frac{\left(kl_--k_-l \right)^{\lambda}}{(k_-+l_-)^{2\lambda}} \tilde{\Phi}(p)\tilde{\bar{\Phi}}(k)\tilde{\bar{\Phi}}(l)
\eea
\ndt The momentum conserving delta function $\delta^4(p+k+l)$ implies that
\begin{eqnarray}
&&\spa{l}.{p}=\sqrt{\frac{2}{p_{-} l_{-}}}\left(kl_{-}-l k_{-}\right)=\frac{\sqrt{k_{-}}}{\sqrt{-\left(k_{-}+l_{-}\right)}}\spa{k}.{l} \\
&&\spa{p}.{k}=\sqrt{\frac{2}{p_{-} k_{-}}}\left(kl_{-}-l k_{-}\right)=\frac{\sqrt{l_{-}}}{\sqrt{-\left(k_{-}+l_{-}\right)}}\spa{k}.{l}\nn\ ,
\end{eqnarray}
\ndt The cubic vertex is then
\bea
\frac{\left(kl_--k_-l \right)^{\lambda}}{(k_-+l_-)^{2\lambda}}&&=\left[\left(\frac{kl_--k_-l}{k_-l_-}\right)(k_-+l_-)\right]^{\lambda}\frac{(k_-l_-)^{\lambda}}{(k_-+l_-)^{3\lambda}}\nn\\
&&=\displaystyle \left[\frac{\langle k l \rangle^{3}}{\langle lp \rangle \langle pk\rangle}\right]^{\lambda}\frac{(k_-l_-)^{\lambda}}{(k_-+l_-)^{3\lambda}}\nn\\
&&=\displaystyle \left[\frac{\langle k l \rangle^{3}}{\langle lp \rangle \langle pk\rangle}\frac{k_-l_-}{(k_-+l_-)^{3}}\right]^{\lambda}={L_3^{\lambda}}_{\mbox {susy}}\ .
\eea
\ndt The light-cone cubic vertex for $\mathcal N=4$ superYang-Mills was previously shown to be~\cite{SA8}
\bea
L_3^{{\mathcal N}=4}=\displaystyle \left[\frac{\langle k l \rangle^{3}}{\langle lp \rangle \langle pk\rangle}\frac{k_-l_-}{(k_-+l_-)^{3}}\right]\ .
\eea
\ndt Therefore the coefficient of the cubic vertex in maximally supersymmetric higher spin theories is equal to the corresponding vertex in the $\mathcal N=4$ theory, raised to the power $\lambda$.
\bea
\label{rel}
{L_3^{\lambda}}_{\mbox {susy}}=\displaystyle \left[\frac{\langle k l \rangle^{3}}{\langle lp \rangle \langle pk\rangle}\frac{k_-l_-}{(k_-+l_-)^{3}}\right]^{\lambda}=\left[L_3^{{\mathcal N}=4} \right]^\lambda\ .
\eea

\noindent Assuming that light-cone higher spin quartic vertices {\it {exist}} and can be written down in this non-covariant gauge, it is likely that the structural relationship in (\ref{rel}) will hold at higher orders as well, suggesting the existence of KLT-like relations~\cite{KLT}.
\vskip 0.3cm

\begin{center}
* ~ * ~ *
\end{center}

\ndt It is surprising that the light-cone Hamiltonians describing Yang-Mills, gravity and higher spin fields all exhibit this quadratic form structure\footnote{The Hamiltonian for the BLG theory is also a quadratic form~\cite{BLG}.}. The light-cone Hamiltonian very likely remains a quadratic form with the $\mathcal W$ picking up higher order corrections. That the form appears only in the pure and maximally supersymmetric versions seems rather striking. There are a number of issues to investigate including implications for higher spin symmetries, residual gauge invariance~\cite{SA1} and particularly field configurations $\mathcal W= 0$ with vanishing energy. 
\vskip 0.3cm
\ndt The Dirac-Feynman path integral generates the quantum action and this quadratic form structure seems to suggest a change of variables from $\Phi$ to $\mathcal W$. This idea is reminiscent of the Nicolai map in Yang-Mills theories~\cite{HN}. Given the ubiquitous nature of $\mathcal N=4$ Yang-Mills theory, a connection between these quadratic form structures, the map~\cite{SA9} and integrability should prove extremely interesting.

\appendix

\section{Superfield identities}

\ndt The following identities based on binomial expansions are useful in our calculations~\cite{BBB2}
\bea
\label{Magic}
\begin{aligned}
\sum_{k=0}^{\lambda-1} (-1)^k&{\lambda-1 \choose k} \left(\bar{\partial}^{\left(\lambda-1-k \right)}\partial_-^k \; {\Phi}\; \partial_{-}^{\left(\lambda-k-1 \right)} \bar{\partial}^k {\Phi} \right)\\
=& \sum_{k=0}^{\lambda-1} (-1)^k{\lambda-1 \choose k} \bar{\partial}^{\left(\lambda-1-k \right)}{\partial_-}^{\left(k \right)}\left({\Phi}\; \partial_{-}^{\left(\lambda-k-1 \right)} \bar{\partial}^k {\Phi} \right)
\end{aligned}
\eea

\subsubsection*{Variant 1}
\bea
\label{magic1}
\begin{aligned}
\sum_{k=0}^{\lambda-1} (-1)^k&{\lambda-1 \choose k}\left(\bar{\partial}^{(\lambda-1-k)}\partial_-^k \; {\Phi} \; \partial_-^{(\lambda-k-1)}\bar{\partial}^k \; \theta^1 \bar{d}_1 \; \Phi \right)\\
=&\sum_{k=0}^{\lambda-1} (-1)^k {\lambda-1 \choose k}\bar{\partial}^{(\lambda-1-k)}\partial_-^k \left({\Phi} \; \partial_-^{(\lambda-k-1)}\bar{\partial}^k \; \theta^1 \bar{d}_1 \; \Phi \right)
\end{aligned}
\eea
This identity is adapted from (\ref{Magic}), with the redefinitions $\Phi_1 \equiv \Phi_1$ and $\theta^1 \bar{d}_1 \Phi_2 \equiv \Phi_2$, allowed because the identity is purely combinatorial in nature.

\subsubsection*{Variant 2}
\bea
\label{magic2}
\begin{aligned}
\sum_{k=0}^{\lambda-1} (-1)^k&{\lambda-1 \choose k}\left(\bar{\partial}^{(\lambda-1-k)}\partial_-^k \; {\Phi} \; \partial_-^{(\lambda-k-1)}\bar{\partial}^{(k+1)} \; \theta^1 \bar{d}_1 \; \Phi \right)\\
=&\sum_{k=0}^{\lambda-1} (-1)^k {\lambda-1 \choose k}\bar{\partial}^{(\lambda-1-k)}\partial_-^k \left({\Phi} \; \partial_-^{(\lambda-k-1)}\bar{\partial}^{(k+1)} \; \theta^1 \bar{d}_1 \; \Phi \right)
\end{aligned}
\eea
Again adapted from (\ref{Magic}), with the redefinitions $\Phi_1 \equiv \Phi_1$ and $\theta^1 \bar{d}_1\, \bar\partial \; \Phi_2 \equiv \Phi_2$.


\begin{thebibliography}{99}
\bibitem{SA1}S. Ananth, L. Brink and M. Mali,  {\it JHEP} {\bf 1508}, 153 (2015), arXiv:1507.01068.\\
S. Ananth, L. Brink, S. Majumdar, M. Mali and N. Shah, {\it JHEP} {\bf 1703}, 169 (2017), arXiv:1702.06261.
\bibitem{SA2}{M. Henneaux, D. Persson and P. Spindel, {\it Living Rev. Rel.} {\bf 11} 01 (2008), arXiv:0710.1818 \\
S. Ananth, L. Brink and S. Majumdar, {\it JHEP} {\bf 1811}, 078 (2018), arXiv:1808.02498.\\
S. Ananth, L. Brink and S. Majumdar, {\it JHEP}, {\bf 1801}, 024 (2018), arXiv:1711.09110.\\
S. Ananth, L. Brink and S. Majumdar, {\it JHEP} {\bf 1603}, 051 (2016), arXiv: 1601.02836.}
\bibitem{BBB1}A. K. H. Bengtsson, I. Bengtsson and L. Brink, {\it Nucl. Phys.} {\bf B 227} (1983) 31-40.\\
A. K. H. Bengtsson, I. Bengtsson and N. Linden, {\it Class. Quant. Grav.} {\bf 4} (1987) 1333.
\bibitem{BBB2}A. K. H. Bengtsson, I. Bengtsson and L. Brink, {\it Nucl. Phys.} {\bf B 227} (1983) 41-49
\bibitem{HE}H. Elvang and Y. Huang, arXiv:1308.1697 (2013).
\bibitem{SM}S. Mandelstam, {\it Nucl. Phys.} {\bf B 213} (1983) 149.
\bibitem{lit}A. Bengtsson, I. Bengtsson and N. Linden, {\it Class. Quant. Grav.} {\bf 4} 1333 (1987).\\
R. R. Metsaev, {\it Nucl. Phys. B} {\bf 759}, 147 (2006), arXiv:hep-th/0512342. \\
R. R. Metsaev, {\it Nucl. Phys. B} {\bf 859}, 13 (2012), arXiv:hep-th/0712.3526.\\
A. Bengtsson, arXiv:1205.6117 (2012).\\
A. Bengtsson, arXiv:1604.01974 (2016). 
\bibitem{SA3}S. Ananth, A. Kar, S. Majumdar and N. Shah, {\it Nuclear Physics B} {\bf 926}, 11 (2017), arXiv:1707.05871.
\bibitem{SA4}S. Ananth, L. Brink, S. Kim and P. Ramond, {\it Nuclear Physics B} {\bf 722}, 166 (2005), arXiv:hep-th/0505234.
\bibitem{SA5}S. Ananth, L. Brink, R. Heise and H. G. Svendsen, {\it Nucl. Phys. B} {\bf 753}, 195 (2006),  arXiv: hep-th/0607019.
\bibitem{SA6}S. Ananth, {\it JHEP} {\bf 1211} (2012), arXiv:1209.4960.\\
Y. Akshay and S. Ananth, {\it J. Phys. A} {\bf 47}, 4, 045401 (2014), arXiv:1304.8082.
\bibitem{SA7}Y. Akshay and S. Ananth {\it Nucl.Phys. B} {\bf 887}, 168 (2014), arXiv:1404.2448.\\
Y. Akshay and S. Ananth, {\it Phys. Rev. D} {\bf 91}, 085029 (2015), arXiv:1504.00967.
\bibitem{SA8}S. Ananth, S. Kovacs and S. Parikh, {\it JHEP} {\bf 05}, 051 (2011), arXiv:1101.3540.
\bibitem{KLT}H. Kawai, D.C. Lewellen and S.H.H. Tye, {\it Nucl. Phys. B} {\bf 269}, 001 (1986).\\
S. Ananth and S. Theisen, {\it Phys. Lett. B} {\bf 652}, 128 (2007), arXiv:0706.1778.\\
S. Ananth, {\it Int. J. Mod. Phys. D} {\bf 19}, 2379 (2010) 2379, arXiv: 1011.3287.
\bibitem{BLG}J. Bagger and N. Lambert, {\it Phys. Rev. D} {\bf 77}, 065008 (2008), arXiv:0711.0955. \\
A. Gustavsson, {\it Nucl. Phys. B} {\bf 811}, 66 (2009), arXiv:0709.1260. \\
D. Belyaev, L. Brink, S. Kim and P. Ramond, {\it JHEP} {\bf 04}, 026 (2010), arXiv:1001.2001.
\bibitem{HN}{H. Nicolai, Nucl. Phys {\bf B176} (1980) 419.\\
K. Dietz and O. Lechtenfeld, Nucl. Phys. {\bf B259} (1985) 397.}
\bibitem{SA9}S. Ananth, H. Nicolai, C. Pandey and S. Pant, {\it J. Phys. A}  {\bf 53, 17}, 174001 (2020), arXiv:2001.02768
\end{thebibliography}
\end{document}